\documentclass[10pt,conference,compsocconf]{IEEEtran}  
\IEEEoverridecommandlockouts

\usepackage{xspace}
\usepackage{amsmath, amssymb}
\usepackage{mathptmx}
\usepackage{nicefrac}
\usepackage{booktabs}
\usepackage{psfrag}
\usepackage{listings}
\usepackage{tikz,tikz-timing,,pgfplots}
\usetikzlibrary{pgfplots.groupplots}
\usepackage{url}

\usepackage{enumitem}
\usepackage{comment}

\usepackage[nolist,nohyperlinks]{acronym}
\acrodef{WSN}[WSN]{Wirless Sensor Network}

\usepackage{tikz,xstring,siunitx,pgfplots,psfrag}
\usepackage{pgfplotstable}
\usepackage{ifthen}

\usetikzlibrary{shapes,arrows}
\usetikzlibrary{calc,patterns,decorations.pathmorphing,decorations.markings}
\usetikzlibrary{positioning,automata}
\usetikzlibrary{pgfplots.groupplots}

\tikzstyle{block} = [draw, rectangle, minimum height=1em, minimum width=1em]
\tikzstyle{sum} = [draw, circle, node distance=1.5cm]
\tikzstyle{input} = [coordinate]
\tikzstyle{output} = [coordinate]

\usepackage{color}

\definecolor{mygreen}{rgb}{0,0.6,0}
\definecolor{mygray}{rgb}{0.5,0.5,0.5}
\definecolor{mymauve}{rgb}{0.58,0,0.82}

\usepackage{listings}
\usepackage[colorinlistoftodos, textwidth=20mm, textsize=footnotesize]{todonotes}

\begin{document}
\onecolumn

\section*{IEEE Copyright Notice}
Copyright (c) 2017 IEEE.
Personal use of this material is permitted. Permission from IEEE must be
obtained for all other uses, in any current or future media, including
reprinting/republishing this material for advertising or promotional purposes,
creating new collective works, for resale or redistribution to servers or lists,
or reuse of any copyrighted component of this work in other works.\\
\\
F. Terraneo, F. Riccardi, A. Leva, "Jitter-compensated VHT and its application to WSN clock synchronization" IEEE Real-Time Systems Symposium (RTSS), Paris, France, December 2017\\
\\
https://doi.org/10.1109/RTSS.2017.00033
\twocolumn
\clearpage

\title{\LARGE \bf Jitter-compensated VHT and its application to WSN clock synchronization
}

\author{\IEEEauthorblockN{Federico Terraneo\IEEEauthorrefmark{1},
                          Fabiano Riccardi\IEEEauthorrefmark{2},
                          Alberto Leva\IEEEauthorrefmark{1}
                         }
        \IEEEauthorblockA{\IEEEauthorrefmark{1}Politecnico di Milano, Italy.
                           Email: \{federico.terraneo,alberto.leva\}@polimi.it             
                         }
        \IEEEauthorblockA{\IEEEauthorrefmark{2}Former graduate student at the Politecnico di Milano
                         }
}

\maketitle
\thispagestyle{empty}
\pagestyle{empty}

\begin{abstract}
Accurate and energy-efficient clock synchronization is an enabler for many applications of Wireless Sensor Networks. A fine-grained synchronization is beneficial both at the system level, for example to favor deterministic radio protocols, and at the application level, when network-wide event timestamping is required. However, there is a tradeoff between the resolution of a WSN node's timekeeping device and its energy consumption. The Virtual High-resolution Timer (VHT) is an innovative solution, that was proposed to overcome this tradeoff. It combines a high-resolution oscillator to a low-power one, turning off the former when not needed. In this paper we improve VHT by first identifying the jitter of the low-power oscillator as the current limit to the technique, and then proposing an enhanced solution that synchronizes the fast and the slow clock, rejecting the said jitter. The improved VHT is also less demanding than the original technique in terms of hardware resources. Experimental results show the achieved advantages in terms of accuracy.

\end{abstract}

\acresetall

\section{Introduction}
\label{sec:Intro}

Precise timing is essential for embedded systems, and in general for computing systems that interact with the physical world. Timing requirements arise out of the need to read sensors and perform actuation at a rate prescribed by the applications, as well as to schedule computations. The design of such systems requires specialized knowledge, and their importance is testified by both a huge research literature and a number of industrial applications.

A computing system's notion of time ultimately comes from hardware counters and event timers, which are used to query the time, and schedule events. As counters are fed by a clock signal, their accuracy depends on that of the clock source.

Inexpensive quartz crystal oscillators have made it possible to provide even cost-sensitive embedded systems with clocks having an accuracy of a few tens of parts per million. This is enough for the requirements of all but the most demanding isolated systems, but for distributed ones, the \emph{scenario} is completely different. When multiple devices need to cooperate, clocks differing by even a few parts per million cause their notion of time to progressively diverge, which very easily turns into an unacceptable situation.

Since distributing a single clock to all devices in a network is resource-consuming, thus hardly feasible especially with wireless connections, it is not surprising that clock synchronization is a very relevant and well studied topic, that dates back to the dawn of computing systems~\cite{bib:Lamport-1978a,bib:Mills-1991a}, but where innovations are still being introduced nowadays~\cite{7509657,bib:AnwarEtAl-2016a}.

Clock synchronization is a fundamental topic for distributed computing systems at large, but in the the Wireless Sensor Network \acp{WSN} domain, it takes a particular flavor. WSNs exhibit at the same time extreme energy constraints and potentially very strict application timing requirements---think for example of sound localization~\cite{bib:MarotiEtAl-2004a,Awaki:2016:SSL:2968219.2971365}. Such needs, coupled with the possibility/opportunity to tailor the entire communication protocol stack, gave rise to unique solutions which are still extending the state of the art in clock synchronization.

One of these solutions is the VHT~\cite{bib:SchmidEtAl-2010b} or Virtual High-resolution Time, which enabled low-energy high-precision timekeeping. VHT is a powerful technique, where two oscillators are used cooperatively: a low-power, low-frequency one is always active to keep the time, and a high-frequency but higher-power one is turned on only when required, giving the impression of an always running high resolution timebase.

In principle, the VHT technique does not set an upper bound to the frequency ratio of the two oscillators, therefore -- again, in principle -- it can be used to increase the resolution of a low-power timekeeping device indefinitely. However, when implementing VHT, we discovered that as the ratio between the clocks increases, the jitter of the low-frequency oscillator limits the accuracy of the overall time representation.
This led us to develop an improved VHT, that instead of combining the two clocks like the original technique, \emph{synchronizes} the high-frequency clock to the slow one.

The contributions of this paper can thus be summarized as follows. First, we provide a model for the VHT technique that allows to study the effect of the nonidealities of the two clocks on the time representation. Then, we propose an improved VHT solution that mitigates the effect of the low-frequency clock jitter, and at the same time is less demanding than the original one in terms of hardware resources. The proposed solution has been implemented on the WandStem WSN node~\cite{bib:TerraneoEtAl-2016a}, and experimental results are reported to show the achieved accuracy improvement.

\section{Related work}
\label{sec:RelWork}

The literature on WSN clock synchronization was historically divided in two broad categories. The first one comprehends works that address clock synchronization with a specific emphasis on low-energy nodes, that have to operate for up to years without replacing batteries. The second one is made of works that aim at pushing the boundaries of WSN synchronization, without considering power consumption.

Typical of the first category is the use of a low-frequency timebase, usually with an inexpensive 32 kHz crystal and  a consumption in the order of $\mu$A, depending on the microcontroller aboard the node. Works in this category include for example DMTS~\cite{bib:PingSu-2003a}, a simple synchronization scheme that periodically overwrites the local clock of each node with a received timestamp, Tiny-Sync~\cite{bib:YoonEtAl-2007a}, that performs skew compensation by trying to constrain the possible skew values using a set of inequalities, FBS~\cite{bib:ChenEtAl-2010a}, that compensates skew using PI control, UTSR~\cite{bib:ChenEtAl-2011a}, that uses broadcast time stations, and hardware assisted clock synchronization~\cite{bib:BuevichEtAl-2013a}, that instead relies on powerline frequency.

Although in this category solutions have been proposed to synchronize within one clock tick, like FLOPSYNC-QACS~\cite{bib:TerraneoEtAl-2016b}, the low resolution of the timebase inevitably results in a synchronization lower bound, that with the typically used 32 kHz crystals is of approximately 30.5$\mu$s.

Works in the second category, that conversely address high-resolution clock synchronization, correspondingly rely on hardware timers clocked by a high-frequency crystal oscillator, usually at a few MHz and above. This category includes RBS~\cite{bib:ElsonEtAl-2002a}, that synchronizes nodes by exchanging local timestamps of a reference packet and compensates skew with linear regression, TPSN~\cite{bib:GaneriwalEtAl-2004a}, that creates a spanning tree of the WSN and then performs pairwise synchronization, FTSP~\cite{bib:MarotiEtAl-2004a}, that introduces flooding as a way to disseminate time, and PulseSync~\cite{6777583}, that improves upon FTSP as for the packet dissemination. The most recent proposals in this category, such as TATS~\cite{bib:LimEtAl-2016a}, also compensate for propagation delays to further improve synchronisztion.

The use of a high-frequency timebase allows these works to achieve microsecond-level synchronization, and in some cases to reach the sub-$\mu$s range. However, such a timebase severely affects the power consumption of the entire node. Also, in most microcontroller architectures, leaving the high-frequency oscillator active prevents entering the deep sleep state, which further increases power consumption.

The introduction of VHT~\cite{bib:SchmidEtAl-2010b} changed the entire \emph{scenario}, allowing for the first time high-resolution and low-power synchronization. Works that adopted a VHT-like timebase, like FLOPSYNC-2~\cite{bib:TerraneoEtAl-2014a}, achieved both sub-$\mu$s clock synchronization and sub-$\mu$A consumption overhead, and extensions such as Reverse Flooding~\cite{bib:TerraneoEtAl-2015a} also added propagation delay compensation. The VHT technique has also been an enabler for protocols requiring tight time synchronization to achieve constructive interference~\cite{7536339}, and for applications such as high-performance sound-based localization~\cite{Awaki:2016:SSL:2968219.2971365}.

For completeness, the power consumption of clocks has been addressed by other works, such as the Tunable Tick Resolution approach~\cite{7103279}, that aims at reducing the interrupt rate for timekeeping in operating systems (OSes) needing a periodic tick also in deep sleep. Fully hardware synchronization solutions have been proposed as well, see e.g. ~\cite{Suzuki2015576}, although the quoted work focuses more on synchronizing the radio wakeup for efficient protocols than at exposing high resolution time to the OS and the applications.

Summarizing after this brief review, VHT is definitely a step forward in low-power timekeeping and a promising technique for a number of applications. In its original form, it has however a few weaknesses as for \emph{accuracy} and resource demand, that in the following we evidence, characterize, and address.

\section{Problem statement and model}

In this section we describe the abstract interface of a generic timekeeping device, show in detail the principle of operation of the original VHT algorithm, and point out the effect of the nonidealities of its timebases.

\subsection{The abstract timekeeping device}

A timekeeping device is a hardware component, coupled with some support software, that is used by an OS to expose timing information to the applications, and to schedule computations. 
This component should expose to the OS at least two fundamental operations, that we name here \emph{get\_time} and \emph{set\_event}, plus optionally two additional ones -- hereinafter \emph{get\_hw\_event\_timestamp} and \emph{set\_hw\_event} --  to facilitate clock synchronization.

\begin{itemize}

\item \emph{get\_time} is an operation that returns the current time. It is the fundamental primitive to allow the OS and the applications to be aware of that time, whatever the purpose of this awareness is.

\item \vspace{1mm}\emph{set\_event} is an operation that causes an event, usually an interrupt service routine, to be called at a given time in the future. The time when the event should occur can be either specified as an absolute time point in the timeline of \emph{get\_time}, or as a relative time interval starting from the call to \emph{set\_event} itself. Its use is primarily to allow the OS scheduler to schedule tasks.

\item \vspace{1mm}\emph{get\_hw\_event\_timestamp} is an optional operation that allows to know the exact time when some hardware-generated event external to the CPU has occurred, for example when a packet has been received through a communication interface. Some hardware implementations provide support for timestamping also application-related events, such as sensor readings, when high timing accuracy is required by applications~\cite{Awaki:2016:SSL:2968219.2971365,bib:TerraneoEtAl-2016a}.

\item \vspace{1mm}\emph{set\_hw\_event} is an optional operation that allows to generate an event external to the CPU, usually a rising edge on some output port, at a given absolute time point in the future. When connected to a communication transceiver, it allows to precisely control the time when a packet is sent without software-induced latencies. Also in this case, additional output channels can be dedicated to applications~\cite{bib:TerraneoEtAl-2016a}.

\end{itemize}

Although the last two primitives are optional, they are becoming common in WSN nodes as their availability allows to increase the accuracy of clock synchronisation~\cite{bib:TerraneoEtAl-2014a,bib:TerraneoEtAl-2015a,bib:LimEtAl-2016a,bib:FerrariEtAl-2011a} especially in multi-hop networks, as they eliminate latencies caused by the software, that accumulate as the number of hops increase. Additionally, also modern clock synchronization protocols over wired network such as PTP~\cite{7509657} do require hardware packet timestamping.

\subsection{A brief review of VHT}

Energy optimization in WSNs is performed by keeping nodes in a low power sleep state as long as possible, with values of 99\% and above being practically achievable~\cite{Istomin:2016:DPS:2994551.2994558}. In this state, the CPU and the radio transceiver are not operational. Only the timekeeper is active, as this is needed to maintain the time, and schedule wakeups. The timekeeper resolution poses a lower bound to the notion of time a node can have, as clock synchronization cannot exceed the hardware tick resolution. However, the timekeeper consumption is a function of its clock frequency, which results in a tradeoff between timing resolution and energy efficiency.

VHT solves this by relying on \emph{two} hardware clocks, a high-frequency one, with frequency $f_h$, that can be turned off during deep sleep periods, and a low-frequency one, with frequency $f_l$. We define the ratio between the two frequencies as
\begin{equation}
\phi_0 = \frac{f_h}{f_l}.
\end{equation}
Most microcontrollers already support such a setting naturally, as the fast clock is the one for the CPU, and the slow one is a 32768 Hz clock for the RTC.

To fully describe the VHT, it is necessary to briefly review how the typical hardware timer in a microcontroller works. Such a device has a hardware counter, that is incremented at the rate of the timer's clock frequency, and one or more hardware units that can be software configured either in \emph{capture mode}, where they store in a register a snapshot of the counter at the occurrence of an external event, or in \emph{compare mode}, where the content of a register is compared against the timer counter, and an event is generated when they are equal. Both in capture and compare mode, it is possible to configure an interrupt to be raised when the event occurs, to allow the software to handle that event.

The VHT paper shows how to implement in software the \emph{get\_hw\_event\_timestamp} operation in Section 4.1 of~\cite{bib:SchmidEtAl-2010b}, and this implementation is what we call ``original VHT'' in this paper.
This implementation uses two hardware timers, clocked respectively by the fast and slow clock. The event source is connected to a capture channel of both timers, to take a timestamp $h_1$ of the high-frequency clock and a timestamp $l_0$ of the low-frequency one every time an event occurs. Either of the two is configured to generate an interrupt, in order to inform the software about the event. Additionally, a second capture channel of the high-frequency timer is connected to the low-frequency clock signal, taking a timestamp $h_0$ of every rising edge of the low-frequency clock. This channel is not configured to generate interrupts.

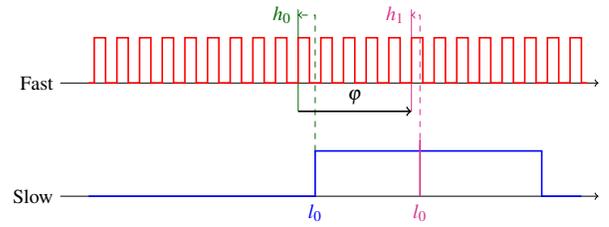
\begin{figure}
 \begin{center}
  \resizebox{0.90\columnwidth}{!}{\begin{tikzpicture}

\draw[->] (-0.5,2) --
     node[pos=0,left] {Fast} (9,2);

\draw[->] (-0.5,0) --
     node[pos=0,left] {Slow} (9,0);

\foreach \x in {0,0.2,...,4.3}
  \draw[red,thick]
       (2*\x,2)--++(0.1,0)--++(0,0.8)
        --++(0.2,0)--++(0,-0.8)--++(0.1,0);


\draw[blue,thick] (0,0)
  -- ( 4,  0) -- ( 4,0.8)
  -- ( 8,0.8) -- ( 8,  0)
  -- ( 8.7,  0);


\draw[->,green!40!black,dashed]
     (4,0.8) -- (4,3.2) --
     node[left,pos=1] {$h_0$} (3.7,3.2);
\draw[green!40!black] 
     (3.7,3.3) -- (3.7,2.8);
\draw[green!40!black] 
     (3.7,2) -- (3.7,1.5);
\node[blue] at (4,-0.3) {$l_0$};

\draw[thick,magenta!90!black]
     (5.85,0) -- (5.85,1);
\draw[->,magenta!90!black,dashed]
     (5.85,0.8) -- (5.85,3.2) --
     node[left,pos=1] {$h_1$} (5.7,3.2);
\draw[magenta!90!black] 
     (5.7,3.3) -- (5.7,2.8);
\draw[magenta!90!black] 
     (5.7,2) -- (5.7,1.5);
\node[magenta!90!black] at (5.85,-0.3)
     {$l_0$};

\draw[->,thick] (3.7,1.5) --
     node[pos=0.5,above] {$\varphi$} (5.7,1.5);


%
%
%
%
%
%

\end{tikzpicture}}
 \end{center}
 \caption{Operation of the original VHT algorithm in the ideal case. Low frequency oscillator ($h_0$) and event ($h_1$) being timestamped by the high frequency clock.}
 \label{fig:VHT-SlowClockJitter}
\end{figure}

When an event occurs, as shown in Figure~\ref{fig:VHT-SlowClockJitter}, the interrupt routine reads $h_0$, $h_1$ $l_0$ and computes the event timestamp as
\begin{equation}
timestamp = l_0 \phi_0 + ((h_1 - h_0) mod \phi_0).
\label{eqn:original-vht-formula}
\end{equation}
The \emph{modulo} operation is necessary because if a rising edge of the slow clock occurs after an event, but before the software routine reads $h_0$, then $h_0$ and $h_1$ will be reversed in the timeline. As can be noticed, the high-frequency clock is only used to measure the time interval between the event and the last edge of the slow clock, identified by $\phi$ in the figure. This means that it can be turned off when not needed, without losing the notion of time. 

\subsection{Issues with the VHT algorithm}
\label{sec:vhtIssues}

The VHT technique allows to achieve the \emph{resolution} of the high-frequency timer without the cost of having it always clocked, but does it allow to achieve also its \emph{accuracy}? To answer this question, it is necessary to introduce clock jitter~\cite{1275150}, a nonideality of crystal oscillators that causes their period to change by a small and random amount at each clock cycle. The point here is that the jitter of the low-frequency clock is small compared to \emph{its} period, but there is no guarantee that it will be small also with respect to the period of the \emph{fast} clock. This introduces a limit to the ratio of the two clocks $\phi_0$, beyond which the VHT accuracy will not increase.
In detail, jitter causes an additive disturbance to the edges of the slow clock, that affects the timestamp $h_0$. Given the nature of~\eqref{eqn:original-vht-formula}, an additive disturbance in $h_0$ will propagate through the difference and the \emph{modulo}, and will finally introduce uncertainty in the event timestamp, degrading accuracy.

To quantify this effect, we performed three experiments.
First, to see the jitter propagation in an ideal setting, we performed a Monte Carlo simulation where 100000 uniformly distributed events over a 100s horizon are timestamped using the original VHT with a 60ns low-frequency clock jitter. The simulation considered a 48MHz high-frequency clock, and a 32768Hz low-frequency one.
Then, to show that the issue exists in real-world hardware, and is not specific to a single particular platform, we implemented the VHT algorithm in two quite different microcontroller boards. The first one is an off-the-shelf stm32f4discovery board. This has a high-frequency clock at 84MHz but lacks a 32kHz crystal, hence we soldered to it an ecliptek EC26 32768Hz one. The second board is a WandStem WSN node, that has a 48MHz high-frequency clock and uses a model CPFB 32kHz crystal by Cardinal Components.

It should be noted that the hardware RTC of both boards lacks an input capture feature, and thus to implement VHT we had to route the output of the 32kHz oscillator to a different timer than the RTC. As only the RTC can function when the microcontroller is in deep sleep, this setup prevents the microcontroller from entering deep sleep, making VHT practically useless but serving the experiment's scope of seeing how the jitter propagates through the algorithm.

\begin{figure}[tb]
 \centerline{\includegraphics[width=0.95\columnwidth]{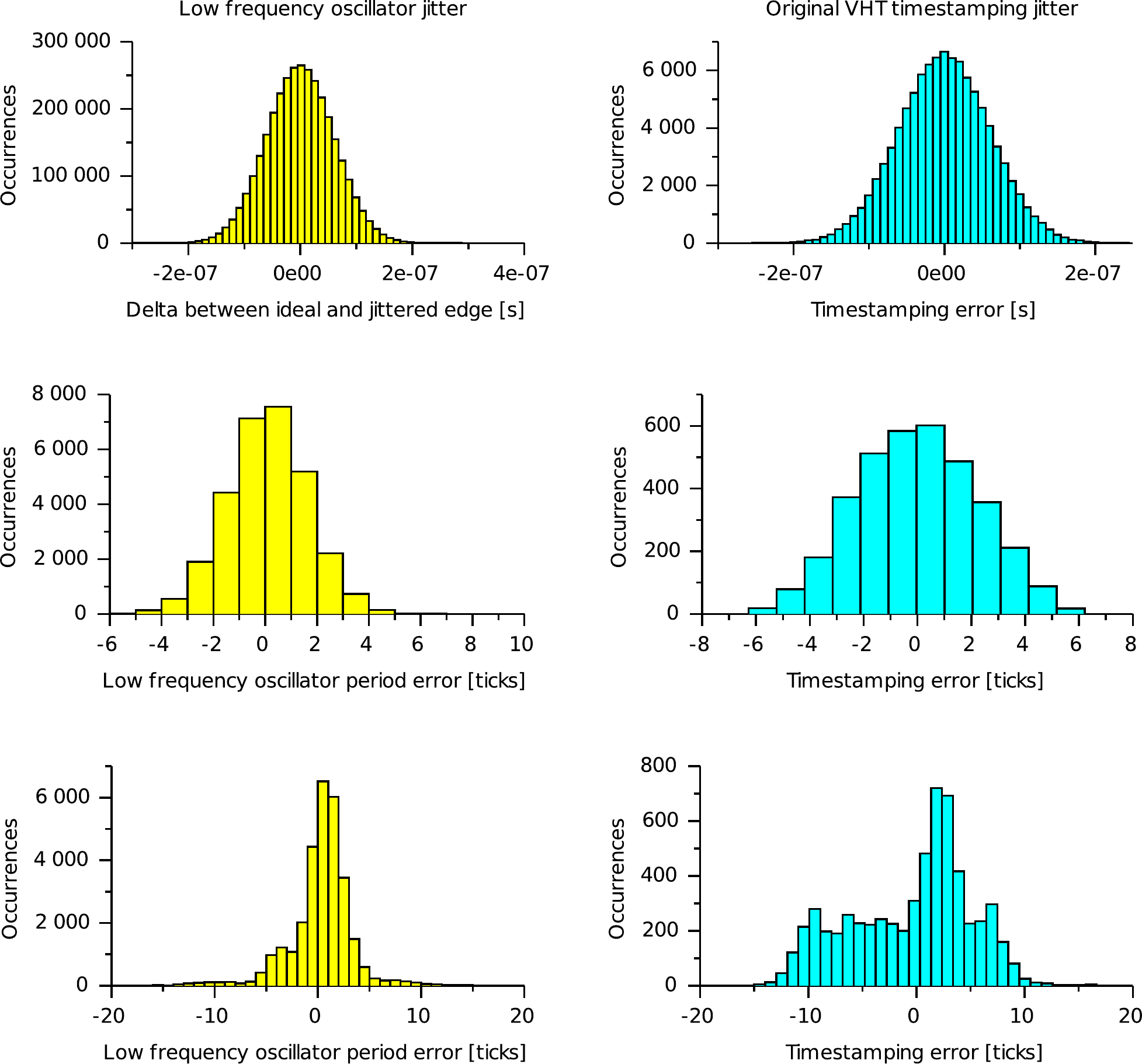}}
 \caption{Low frequency clock jitter vs original VHT algorithm timestamping jitter. Top row, Monte Carlo simulation; middle row, stm32f4discovery; bottom row, WandStem WSN node.}
 \label{fig:vht-jitter-comparison}
\end{figure}

Figure~\ref{fig:vht-jitter-comparison} shows the result of the experiments. The left column shows the jitter of the low-frequency oscillator as timestamped by the high-frequency one. The right column shows the jitter of timestamps taken using the original VHT algorithm.
In all three implementations we observed some timestamps with an error of 30.5$\mu$s, but these were caused by a race condition that will be described later on in this section, so we removed them to focus only on jitter.

The top row shows the results for the Monte Carlo simulation. In this case the jitter of the VHT timestamping is 60.4ns, a value remarkably close to the 60ns jitter of the low-frequency clock. Thus, in an ideal setting where the only disturbance is jitter and quantization noise, the standard deviation of VHT equals the jitter of the slow clock.
The middle row shows the results for the stm32f4discovery. In this case the jitter of the low-frequency clock had a standard deviation of 19ns, and the maximum error was $\pm$ 7.5 ticks of the fast clock. In this board the fast clock is obtained through a PLL, thus the high-frequency clock is not fully stable, and exhibits a nonideality called wander~\cite{bib:pllwander} (a slow random drift in the timescale of seconds). After removing the wander by means of a high pass filter, the jitter of the VHT algorithm is as shown in the middle right plot of Figure~\ref{fig:vht-jitter-comparison}, and the standard deviation is 27ns.
With the WandStem board, the jitter of the low-frequency clock had a standard deviation of 63ns, and the maximum error was $\pm$ 17.5 ticks of the fast clock. The VHT jitter was instead 116ns; the probability distribution spreads across more values due to temperature changes during the experiment, introducing time-varying skew.

As can be seen, in all three experiments the jitter of VHT is never less than the jitter of the underlying slow clock, showing how the VHT timestamping algorithm does not attenuate the low-frequency clock jitter. Moreover, in both hardware implementations, the jitter of the slow clock is several ticks of the fast one, supporting the statement that slow clock jitter is a limiting factor to accuracy.

\begin{figure}[tb]
 \centerline{\includegraphics[width=0.95\columnwidth]{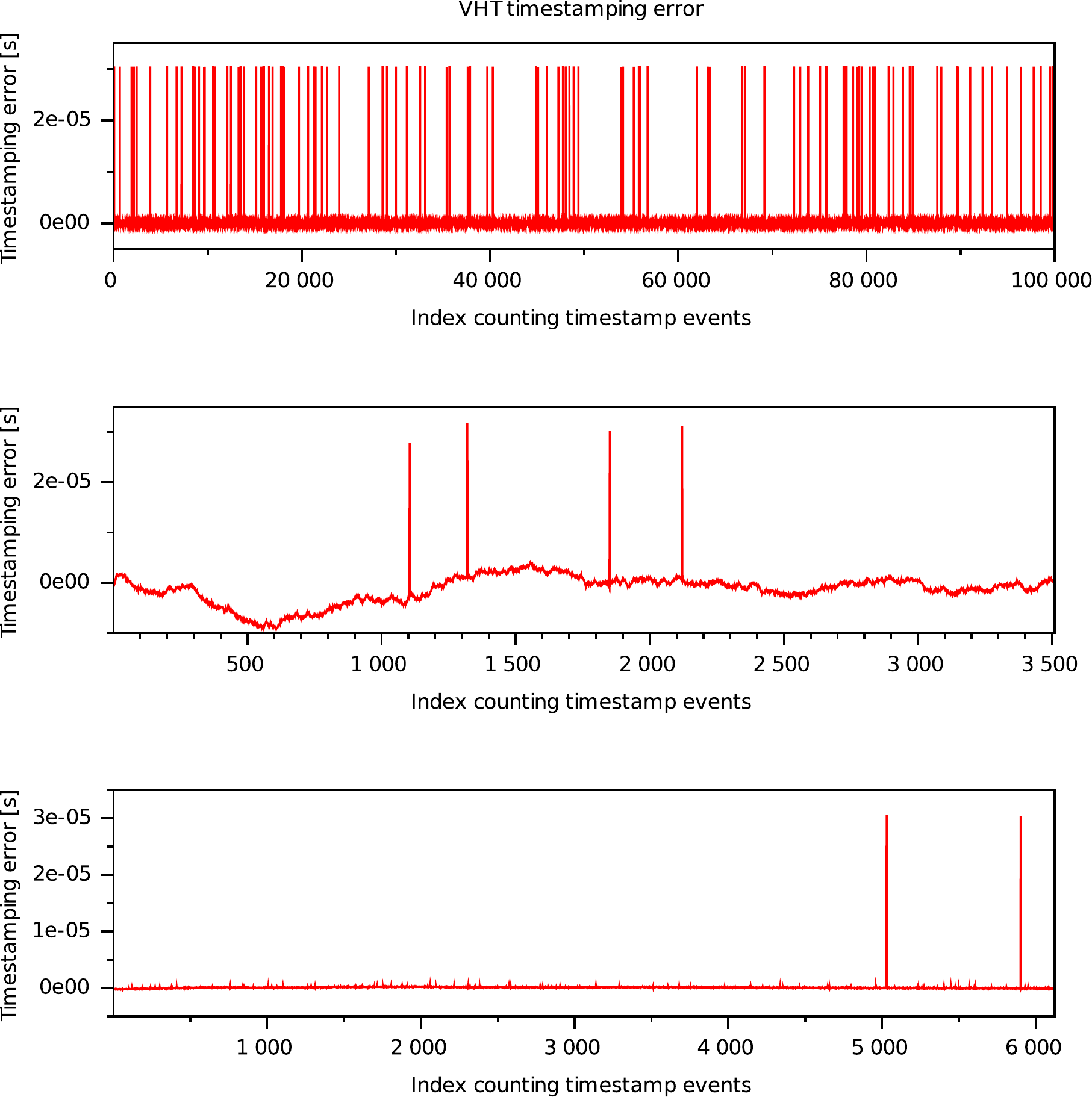}}
 \caption{VHT algorithm timestamping error including the timestamps where a race condition occurs. Top: Monte Carlo simulation, middle: stm32f4discovery implementation, bottom: WandStem implementation. The low-frequency random drift in the middle plot is the wander of the STM32 PLL.}
 \label{fig:vht-error-comparison}
\end{figure}

Figure~\ref{fig:vht-error-comparison} shows instead the raw timestamping error using the original VHT algorithm, in the three experiments above.
Note the peaks of 30.5$\mu$s, which correspond to one tick of the low-frequency clock. These errors are caused by a race condition in the original VHT algorithm.

\begin{figure}[tb]
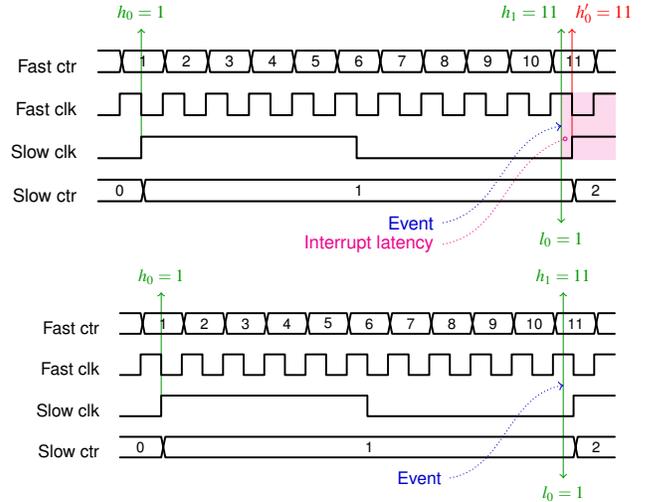

 \centerline{\resizebox{1.00\columnwidth}{!}{\input{./draw/RaceCondition-2.tex}}}
 \centerline{\resizebox{0.92\columnwidth}{!}{\input{./draw/RaceCondition-1.tex}}}
 \caption{Timing diagram of the fast and slow clock showing the VHT algorithm race condition.}
 \label{fig:racecondition}
\end{figure}

To explain this condition, first consider the top timing diagram of Figure~\ref{fig:racecondition}. In this example the ratio $\phi_0$ between the fast and slow clock is 10. Now consider that the phase between the two clocks is such that there is a small time where the fast clock has incremented $\phi_0$ times since the last rising edge of the slow clock. If an event occurs in that time instant, the input capture units will log the following timestamps: $h_1$=11, $l_0$=1. However, since the software interrupt has a latency, the $h_0$ timestamp will be overwritten as soon as the second edge of the slow clock occurs, so instead of $h_0$=1 we will have $h'_0$=11. Applying Equation~\eqref{eqn:original-vht-formula} the wrong timestamp 10 is computed instead of the correct result which is 20.
The bottom part of the figure shows that the race condition occurs even when considering zero interrupt latency, as in this case we will have $h_0$=1 and $h_1$=11, but 10 $mod$ 10 is 0.

\section{Jitter-compensated VHT}

As shown in the previous section, the VHT technique exhibits three issues. The first one is the degradation of the time accuracy as the ratio between the clocks increases, due to the jitter of the low-frequency clock.
The second one is the race condition just shown.
The third one, perhaps less obvious and on which we shall return later on, is the need for two input capture units for each event to be timestamped, which makes its implementation heavy on hardware resources.
The issues above are exacerbated if one wants to use the VHT as an OS timer, because doing so requires to implement the entire timekeeper interface, not just the \emph{get\_hw\_event\_timestamp} operation---a matter out of the scope of the original VHT paper, however.

In this section, we overcome these issues by introducing an improved VHT implementation, that we call \emph{jitter-compensated} VHT. This solutions casts the VHT problem in the clock synchronization framework, and works by \emph{synchronizing} the fast clock to the slow one, by means of the FLOPSYNC-2 technique~\cite{bib:TerraneoEtAl-2014a}, every time the node gets out of deep sleep.
To explain the operation of our jitter-compensated VHT, we proceed in two steps. First we show a na\"{i}ve approach, that only corrects for offset, as a preliminary step to introducing the complete scheme, that also performs skew correction.

\subsection{The na\"{i}ve jitter-compensated VHT}
\label{sec:naiveSolution}

The simplest solution to solve the scalability issues mentioned above consists in using the low-frequency clock to resynchronize the high-frequency one only upon exiting deep sleep, and then perform time-related operations using only the fast clock.
Since all operations are performed using only one timer, this solution is also inherently immune from the race condition described in Section~\ref{sec:vhtIssues}.
To do so, one can measure the offset every time the node goes out of deep sleep, by taking a timestamp $h_0$ of the fast clock corresponding to an edge $l_0$ of the slow clock, as

\begin{equation}
 offset = l_0 \phi_0 - h_0.
\end{equation}

This offset computation can be repeated multiple times, for subsequent edges of the slow clock, averaging the results to reduce the effect of jitter that affects $h_0$. Such an operation can be considered acceptable because the wakeup is an infrequent event, thus the expense in terms of timestamp measurements and average computation, which on a typical implementation can be assumed to be less than 500$\mu$s, does not impact the node performance to a relevant extent.

From then on, timestamps can be taken using only a capture channel of the high-speed timer, and converted to the VHT timeline by simply adding the offset, thus implementing the \emph{get\_hw\_event\_timestamp} operation. The \emph{get\_time} operation can likewise be implemented by reading the high-frequency timer counter and adding the offset, while the \emph{set\_event} and \emph{set\_hw\_event} can be implemented by using a single compare channel of the fast clock timer, which is set to the desired time minus the offset.

This solution clearly solves the hardware resource issue, as all timekeeping operations are implementable with at most one capture/compare channel of the fast clock per operation, plus an additional one to compute the offset at every wakeup from deep sleep.

This solution also solves the clock jitter issue. As timestamps are taken using only the high-frequency clock, the jitter of the low-frequency clock no longer affects each time measurement. For what concerns interval measurements, those obtained by subtracting two timestamps taken while the node is active are entirely taken by the high-frequency timer, and are thus insensitive to the low-frequency clock jitter. If on the contrary the node went to deep sleep in between two timestamps, the jitter effect is significantly mitigated by the averaging in the offset computation.

However, the na\"{i}ve jitter-compensated VHT described so far would not work in practice, due to another nonideality of crystal oscillators, that is, clock skew.
In fact, both the high-frequency and the low-frequency crystal oscillators are affected by skew, and the skews of the two can differ significantly from one another, both in magnitude and as for their dependence on temperature.

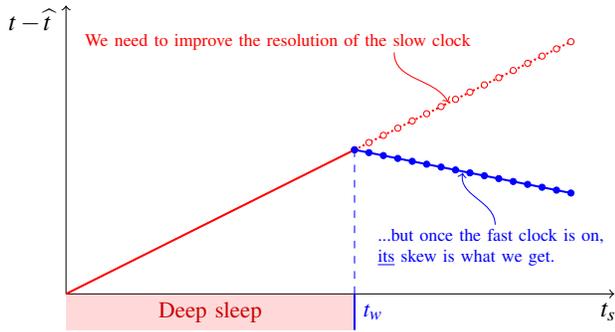
\begin{figure}[t]
 \begin{center}
  \resizebox{0.95\columnwidth}{!}{\begin{tikzpicture}

\draw[->] (0,0) --
     node[pos=0.99,below] {$t_s$} (7.6,0);
\draw[->] (0,0) --
     node[pos=0.95,left,xshift=-2]
     {$t-\widehat{t}$} (0,4);

\fill[draw=none,fill=red,opacity=0.15] 
     (0,0) rectangle (4,-0.5);
\node[red!80!black] at(2,-0.25)
     {\small{Deep sleep}};

\draw[red,thick] (0,0) -- (4,2);
\draw[red,thick,densely dotted] (4,2) -- (7,3.5);
\draw[blue,thick] (4,2) -- (7,1.4);

\foreach \x in {4,4.2,...,7}
 \draw(\x,\x/2)
  node[circle,draw=red,fill=white,
       inner sep=0.3mm] {};
\foreach \x in {4,4.2,...,7}
 \draw(\x,2.8-0.2*\x)
  node[circle,draw=blue,fill=blue,
       inner sep=0.3mm] {};

\draw[blue,thick] (4,-0.5) --
     node[pos=0.5,right]
     {\small{$t_w$}} (4,0);
\draw[blue,dashed] (4,0) -- (4,2);

\node[right,red] at (0.14,3.5)
  {\scriptsize{We need to improve the resolution
   of the slow clock}};
\draw[->,red] (4.55,3.35) to [out=-90,in=120] (5.3,2.66);

\node[right,blue] at (4.2,0.8)
  {\scriptsize{...but once the fast clock is on,}};
\node[right,blue] at (4.2,0.5)
  {\scriptsize{\underline{its} skew is what we get.}};
\draw[->,blue] (5.95,0.96) to [out=90,in=-100] (5.5,1.68);

\end{tikzpicture}

}
 \end{center}
 \caption{Clock skew issue in the na\"{i}ve jitter-compensated VHT}
 \label{fig:VHT-InternalSync-problem}
\end{figure}

Setting without loss of generality the time origin of the low-frequency clock to when the node was first turned on, and denoting by $t_w$ the generic wakeup time, the time reported by the na\"{i}ve jitter-compensated VHT will be
\begin{equation}
C_{naive}(t) = \phi_0 \left\lfloor \int_0^{t_w} f_l + \delta_{fl}(\tau)d\tau \right\rfloor
            +         \left\lfloor \int_{t_w}^t f_h + \delta_{fh}(\tau)d\tau \right\rfloor
\end{equation}
where $\delta_{fl}(t)$ and $\delta_{fh}(t)$ are terms representing the frequency error of the two clocks due to their nonidealities. The first integral is the skew accumulation of the slow clock till the node wakeup, which coincides with the offset computation, while the second integral is the skew accumulation of the fast clock from the wakeup instant onwards. The floor operators account for the tick quantization. Figure~\ref{fig:VHT-InternalSync-problem} shows a graphical representation of the issue.

Synchronization schemes that can correct for skew using the time information received via the radio are well known, but these schemes are meant to compensate a \emph{single} clock skew, not for a \emph{blend} of two skews that depends on how long the node has been out of deep sleep.

To overcome this issue, we propose a hierarchical solution that introduces a local clock skew correction to make the skew of the fast clock equal to that of the slow clock, thereby allowing for the abstraction of an always running, jitter-free, low-power and high-resolution clock with a single skew value.
Once this abstraction is built, an ordinary clock synchronization scheme can compensate the skew of the jitter-compensated VHT, needing no awareness that the clock to synchronize is built with two different crystals.

\subsection{Skew correction in the jitter-compensated VHT}
\label{sec:skewCorrection}

Performing skew correction means synchronizing the fast clock to the slow one. There are various phenomena that make the two clocks diverge, like crystal imperfections, aging, jitter, and temperature variations. These phenomena collectively manifest themselves in the synchronization error, but cannot be measured individually. In problems like this, feedback control is a natural solution, as it permits to contrast error-generating phenomena (in control terms, \emph{disturbances}) based only on the measurement of the error.

We thus address the problem with the FLOPSYNC-2 synchronization framework~\cite{bib:TerraneoEtAl-2014a}, which is based on a feedback controller. However, the frequency content of the disturbances encountered in this work is not the same as in~\cite{bib:TerraneoEtAl-2014a}, and although the feedback control structure can be re-used, the computation of the control signal needs modifying. In this section we illustrate the control problem, review the FLOPSYNC-2 solution, and describe the synthesis of the new control law. 

\begin{figure}[htb]
 \begin{center}
  \resizebox{0.95\columnwidth}{!}{\begin{tikzpicture}

\draw[->] (0,0) --
     node[pos=0.99,below] {$t_s$} (7.6,0);
\draw[->] (0,0) --
     node[pos=0.95,left,xshift=-2]
     {$t-\widehat{t}$} (0,4);

\fill[draw=none,fill=red,opacity=0.15] 
     (0,0) rectangle (2,-0.5);
\node[red!80!black] at(1,-0.25)
     {\small{Deep sleep}};
\draw[blue,thick] (2,-0.5) --
     node[pos=0.5,right]
     {\small{$t_w$}}  (2,0);
\draw[blue,dashed] (2,0) -- (2,1);

\draw[magenta] (2,1) -- (2,3.2);
\draw[magenta] (3,1.5) -- (3,3.2);
\draw[<->,magenta] (2,3) -- (3,3);
\node[left,magenta] at (2,3)
     {\scriptsize{Fast to slow}};
\node[left,magenta] at (2,2.75)
     {\scriptsize{sync period}};

\draw[red,thick] (0,0) -- (2,1);
\draw[red,dashed] (2,1) -- (7,3.5);

\foreach \x in {2,3,...,7}
 \draw(\x,\x/2)
  node[circle,draw=red,fill=white,
       inner sep=0.4mm] {};
\foreach \x in {2.5,3.5,...,6.5}
 \draw(\x,\x/2)
  node[circle,draw=red,fill=white,
       inner sep=0.2mm] {};

\draw[red] (2,1) --(2,2.3);
\draw[red] (2.5,1.25) --(2.5,2.3);
\draw[<->,red] (2,2.1) -- (2.5,2.1);
\node[left,red] at (2,2.1)
     {\scriptsize{Slow period}};

\draw[blue] (2,1) --(2,1.8);
\draw[blue] (2.24,0.9) --(2.24,1.8);
\draw[<->,blue] (2,1.6) -- (2.24,1.6);
\node[left,blue] at (2,1.6)
     {\scriptsize{Fast period}};

\draw(2,1) node[circle,draw=blue,fill=blue,
                inner sep=0.4mm] {};
\draw[blue,thick] (2,1) -- (3.2,0.7);
\foreach \x in {2,2.24,...,3.2}
 \draw[blue,thick](\x,1.5-0.25*\x-0.1) 
       --(\x,1.5-0.25*\x+0.1);

\draw[red] (3,1.5) -- (3,-1.25);
\node[red,left,xshift=2.5] at(3,-0.9)
     {\scriptsize{Expected sync period}};
\node[red,left,xshift=2.5] at(3,-1.15)
     {\scriptsize{in fast clock}};

\draw(3.2,0.7)  
  node[circle,draw=blue,fill=blue,
                inner sep=0.4mm] {};

\draw[blue] (3.2,0.7) -- (3.2,-1.95);
\node[blue,left,xshift=2.5] at(3.2,-1.6)
     {\scriptsize{Actual sync period}};
\node[blue,left,xshift=2.5] at(3.2,-1.85)
     {\scriptsize{in fast clock}};

\fill[draw=none,fill=green!50!black,
      opacity=0.6] 
     (3,0) rectangle (3.2,0.25);
\draw[green!50!black,]
     (3.1,0.125) --
     node[pos=1,right]
     {\scriptsize{Fast to slow sync error}}
     (4,-1);

\draw[blue,thick] (3.2,0.7) -- (4.1,1.0);
\draw(4.1,1,0)  
  node[circle,draw=blue,fill=blue,
                inner sep=0.4mm] {};

\draw[blue,thick] (4.1,1.0) -- (4.95,1.8);
\draw(4.95,1.8)  
  node[circle,draw=blue,fill=blue,
                inner sep=0.4mm] {};

\draw[blue,thick] (4.95,1.8) -- (6.05,2.7);
\draw(6.05,2.7)  
  node[circle,draw=blue,fill=blue,
                inner sep=0.4mm] {};

\draw[blue,thick] (6.05,2.7) -- (7.05,3.35);
\draw(7.05,3.35)  
  node[circle,draw=blue,fill=blue,
                inner sep=0.4mm] {};

\draw[->,thick,green!50!black]
      (5.5,-0.8) 
      to [out=110,in=-60] (3.65,0.85);

\node[right,green!50!black] at (4,0.6)
     {\scriptsize{Fast-slow skew compensation by}};
\node[right,green!50!black] at (4.4,0.3)
     {\scriptsize{the FLOPSYNC-2 technique}};

\end{tikzpicture}}
 \end{center}
 \caption{The FLOPSYNC-2 clock synchronization framework applied to intra-node skew correction.}
 \label{fig:VHT-InternalSync-propSolution}
\end{figure}
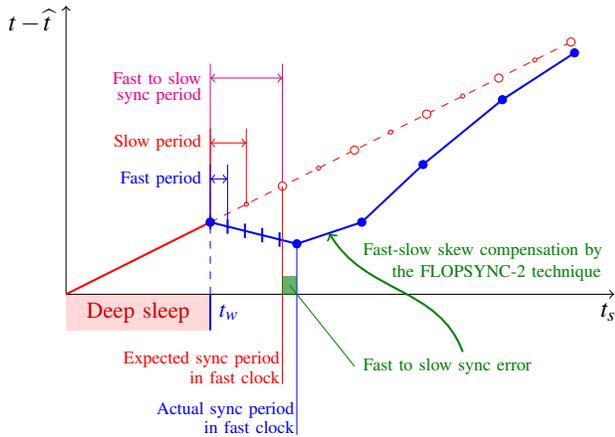

For the physical reasons sketched above and detailed in~\cite{bib:TerraneoEtAl-2014a}, the frequency $f_h$ of the fast clock varies over time with respect to its nominal value $\phi_0 f_{l0}$, where $f_{l0}$ is the nominal frequency of the slow clock. Since we want the fast clock to synchronize to the slow one -- or equivalently, we want the fast clock to be the \emph{slave} and the slow the \emph{master} -- we take the time counted by the latter as the reference. Denoting the fast and slow clock times respectively with $t_h$ and $t_l$, and indicating with $e_{hl}$ the synchronization error, we can write
\begin{equation}
 t_h(t_l) = \int_{0}^{t_l} \frac{\phi_0 f_{l0} + \delta f_h(\tau)}{\phi_0 f_{l0}} d\tau
\end{equation}
whence
\begin{equation}
 e_{hl}(t_l) := t_l-t_h(t_l) = - \int_{0}^{t_l} \frac{\delta f_h(\tau)}{\phi_0 f_{l0}} d\tau.
\end{equation}

As in FLOPSYNC-2, we assume synchronizations to take place at a fixed period, $T_{hl}$ to name it, dictated by the master clock. The error is measured -- see Figure~\ref{fig:VHT-InternalSync-propSolution} -- as the expected minus the actual end of the synchronization period counted in the slave clock. This gives the controlled system exactly the same structure as in~\cite{bib:TerraneoEtAl-2014a}, with the inter-node frequency fluctuation replaced by the intra-node one.

The next expected end of the synchronization period is computed as the previous one plus $T_{hl}$ plus a correction $u_{hl}(k)$, that takes the role of the control signal. Denoting by $e_{hl}(k)$ the error at the $k$-th intra-node synchronization event, computations omitted for brevity lead to write
\begin{equation}
 e_{hl}(k) = e_{hl}(k-1) + u_{hl} (k-1) + d_{hl}(k-1),
 \label{eqn:P}
\end{equation}
where
\begin{equation}
 d_{hl}(k-1) := -\int_{(k-1)T_{hl}}^{kT_{hl}} \frac{\delta f_h(\tau)}{\phi_0 f_{l0}} d\tau
\end{equation}
is the disturbance provided by the fast/slow relative skew cumulated over one synchronization period.

Should this disturbance -- hence the skew -- be constant, non control-centric synchronization schemes (such as those based on regression) could fit. However the skew is not constant at all, and depends on phenomena (aging, thermal stress, jitter) with different time scales. This motivates the choice of a feedback control law designed specifically to counteract the frequency components of that disturbance that are most relevant for the synchronization need at hand.

In~\cite{bib:TerraneoEtAl-2014a}, FLOPSYNC-2 was used to synchronize clocks aboard different nodes, and the most relevant source of disturbance was thermal stress. In the case considered herein the clocks are on the same node, and given the much faster time scale, the main concern is high-frequency jitter. We now proceed to the design of a control law with the specific aim of rejecting high-frequency disturbance components.

\begin{figure}[htb]
 \begin{center}
  \vspace{2mm}\resizebox{0.95\columnwidth}{!}{\begin{tikzpicture}

\draw[->] (0,0)
   -- node[pos=-0.02] {0}
      node[pos=0.99,below] {$\omega$} (10,0);
\draw[->] (0,-2.2)
   -- node[pos=0.97,left] {dB} (0,2.2);

\draw[red,ultra thick] (0.8,2) -- (2.2,0.8)
   -- (7.7,-0.8) -- (9.2,-2);
\node[red] at (1.8,2) {$|L(j\omega)|$};
\draw[red,] (0.8,2) -- (47/15,0)
     -- node[pos=1,below]
     {\Large{$\frac{\omega_c}{\sqrt{\alpha}}$}}
     (47/15,-0.5);
\draw[red,thick,densely dotted] (2.2,0.8)
  -- node[pos=1.0,below]
    {\Large{$\frac{\omega_c}{\alpha}$}}
    (2.2,0);
\draw[red,thick,densely dotted] (7.7,-0.8)
  -- node[pos=1.0,above] {$\beta\,\omega_c$}
     (7.7,0);

\draw[red,] (5,-0.5)
  -- node[pos=0,below] {$\omega_c$}
     (5,0);

\end{tikzpicture}}
 \end{center}
 \caption{Control synthesis: the obtained loop frequency response magnitude.}
 \label{fig:ControlSynthesis-CT-Ljw}
\end{figure}
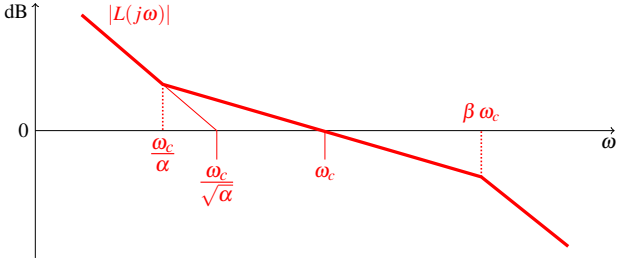

We carry out the design in the continuous time domain, mainly because doing so allows to operate with frequency values directly, and independently of the adopted synchronization period. Putting model~\eqref{eqn:P} in transfer function form and transforming to continuous time, we get 
\begin{equation}
 E_{hl}(s) = P(s) \left( U_{hl}(s)+D_{hl}(s) \right), \quad P(s) = \frac{1}{T_{hl}s},
 \label{eqn:Ps}
\end{equation}
where $s$ is the Laplace transform complex variable. We use an integral controller augmented with a zero-pole pair; this guarantees a 40 dB/decade high-frequency disturbance-to-output roll-off, and allows to prescribe the cosed-loop stability degree. In detail, writing the controller as
\begin{equation}
 C(s)   = \frac{\omega_c^2 T_{hl}}{\alpha s}\,
          \frac{1+s\frac{\alpha}{\omega_c}}{1+s\frac{1}{\beta\omega_c}}, \qquad
 \alpha,\beta > 1,
 \label{eqn:Cs}
\end{equation}
the Bode magnitude plot of the frequency response of the loop transfer function $L(j\omega)=C(j\omega)P(j\omega)$ takes the form of Figure~\ref{fig:ControlSynthesis-CT-Ljw}.

The cutoff frequency $\omega_c$ dictates the closed-loop response speed, while parameter $\alpha$ governs the stability degree, because the phase margin is
\begin{equation}
 \varphi_m = \arctan \left( \alpha \right) - \arctan \left( 1/\beta \right).
 \label{eqn:pm}
\end{equation}
\vspace{2mm}\begin{figure}[htb]
 \begin{center}
  \includegraphics[width=\columnwidth]{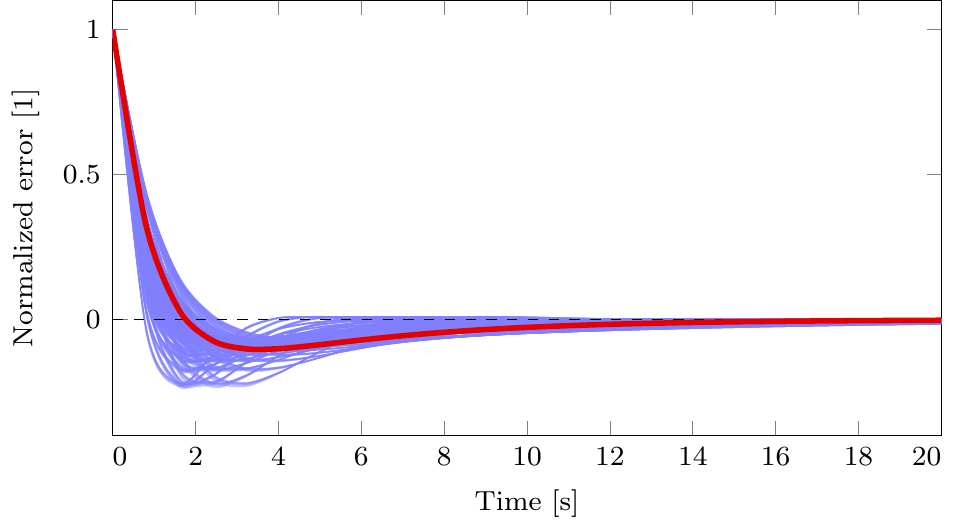}
 \end{center}
 \caption{Sample of the design space exploration conducted to synthesize the controller~\eqref{eqn:Cs}.}
 \label{fig:DSE}
\end{figure}

To determine the parameters in controller~\eqref{eqn:Cs}, we set $T_{hl}$ to the recommended value (see Section~\ref{sec:syncperiodselection}) of 200 ms and carried out a design exploration in the $(\omega_c,\alpha,\beta)$ space, operating in simulation with the process model~\eqref{eqn:Ps}. The goal was to obtain a fast error convergence to zero, while maintaining a good high-frequency disturbance rejection, and an adequate stability degree. A sample of the said exploration, illustrating the error convergence speed, is shown in Figure~\ref{fig:DSE}. After the exploration we chose $\omega_c=5/4$ r/s, $\alpha=25/4$ and $\beta=16$, which correspond in Figure~\ref{fig:DSE} to the red thick response. These values yield a phase margin of about 77$^{\circ}$; a faster settling would result in a lower stability degree, and most important, in a less effective jitter rejection. Applying the backward Euler method, we finally got the discrete-time controller
\begin{equation}
 C(z) = \frac{26z^2-25z}{125z^2-150z+25}.
\end{equation}

This controller settles to compensate the skew within 0.1\% in about 14 s, see again the red thick response in Figure~\ref{fig:DSE}. This is therefore the time it takes when the node is first turned on to correct for 99.9\% of the initial (unknown) skew, which we found to be an adequate performance. Listing~\ref{lst:flopsyncController} reports the implementation of the proposed controller in C++, evidencing its simplicity.

\begin{figure}[tb]
  \lstinputlisting[language=C++,%
  caption={The proposed intra-node clock skew correction controller.},%
  label={lst:flopsyncController}]%
  {./sections/flopsync_controller.cpp}
\end{figure} 

\subsection{Synchronization period selection}
\label{sec:syncperiodselection}

In the case of the inter-node synchronization it is important to keep the synchronization period as long as possible, in order to minimize the radio usage. In the case of intra-node synchronization it is instead beneficial to keep the period as short as possible, because the high-frequency clock is not running while the node is in deep sleep, and thus intra-node synchronization is only possible while the node is active.

The lower limit of the synchronization period is due to the quantization induced by the fast clock resolution
\begin{equation}
q_e = \frac{10^6}{f_h T_{hl}}
\end{equation}
where $q_e$ is the quantization error in parts per million.
Inverting this formula, to synchronize a 48MHz fast clock to within 0.1 ppm, the minimum synchronization period is around 200ms, which can be reduced to 20ms if 1ppm accuracy is enough.

\subsection{The complete jitter-compensated VHT}

To summarize, the complete jitter-compensated VHT works as follows. When the node is powered up, both oscillators are turned on, and the offset between the two clocks is computed as in Section~\ref{sec:naiveSolution}. After that, the skew correction starts as per Section~\ref{sec:skewCorrection}. The node is forced to stay active long enough for the skew correction controller to settle, after which it is allowed to go in deep sleep. Note that during this time the only CPU load caused by our VHT implementation is one interrupt for each synchronization period $T_{hl}$, so the CPU is essentially idle and this time can be proficiently used for other tasks, such as MAC neighbor discovery, and to wait for the first synchronization packet as required by inter-node synchronization schemes.

Every time the node gets out of deep sleep, the offset correction is performed again. Additionally, the last skew correction is preserved when entering deep sleep, so as to apply it upon wakeup. This eliminates the need to wait for the controller to settle every time a wakeup from deep sleep occurs.

As long as the node remains active, the skew correction controller is executed at period $T_{hl}$. Every time the node wakes up to receive the synchronization packet and to perform the inter-node synchronization, it is forced to stay active for at least $T_{hl}$, and thus perform one inter-node skew correction update. This is required to ensure a minimum rate of skew correction updates even with applications that always wake up for smaller periods than $T_{hl}$. Note that in our implementation we decided to tie the minimum rate of the intra-node skew correction to the inter-node synchronization period mainly for convenience, but other options are possible.

The solution has an overhead in terms of hardware resources of two capture/compare units: one output compare unit of the slow clock connected to an input capture unit of the fast clock. These two hardware resources are set up to generate an event every rising edge of the slow clock during the averaging of the offset when waking up from deep sleep, and then reconfigured to generate one event every intra-node synchronization period $T_{hl}$, as long as the node is active, to perform skew correction. All the operations of the timekeeper interface can then be implemented using just one capture/compare unit of the fast clock, except for \emph{get\_time} which requires none.

\section{Evaluation}

The proposed solution was evaluated in terms of lower hardware resource usage, jitter attenuation capability and power consumption.

\subsection{Hardware requirements}

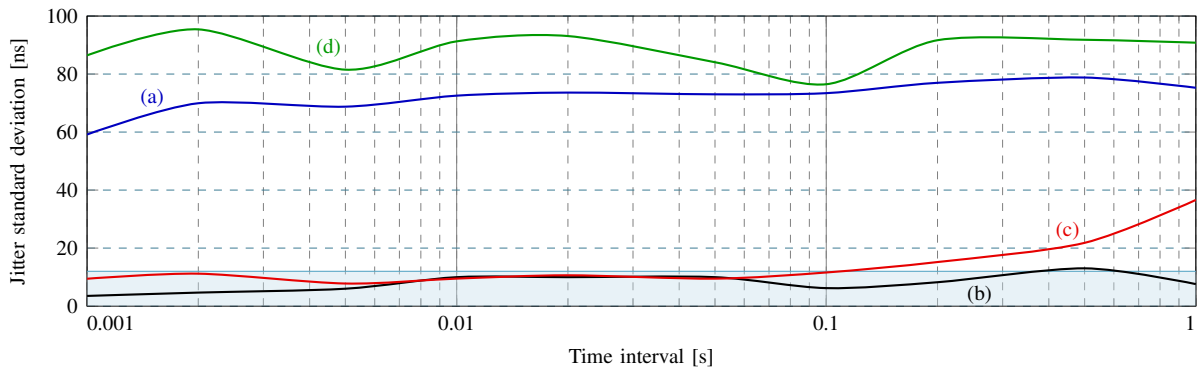
\begin{figure*}[tb]
 \begin{center}
  \begin{tikzpicture}
\pgfplotsset{width=0.90\textwidth,height=0.30\textwidth,
            compat=newest,
            major x grid style={very thin,gray},
            minor x grid style={very thin,dashed,gray},
            major y grid style={very thin,dashed,cyan!50!black},
}

\begin{axis}
[
grid=both,
xmode=log,
log ticks with fixed point,
xmin=0.001,
xmax=1,
xtick={0.001,0.01,0.1,1},
xticklabels={\rlap{0.001},0.01,0.1,\llap{1}},
ymin=0,
ymax=100,
enlargelimits=false,
xlabel={Time interval [s]},
ylabel={Jitter standard deviation [ns]},
font=\footnotesize
]

\fill [fill=cyan!50!gray!30!white,opacity=0.4] (axis cs:0.001,0) rectangle (axis cs:1,12);

\addplot[thin,color=cyan!50!gray] coordinates {(0.001,12)(1,12)};

\addplot[thick,color=blue!75!black,smooth] table [x index=0,y index=1,col sep=comma]
        {./img/jitter-comp-results-withVHT2.csv};
\node[blue!75!black] at (axis cs: 0.0015,72) {(a)};

\addplot[thick,color=black,smooth] table [x index=0,y index=2,col sep=comma]
        {./img/jitter-comp-results-withVHT2.csv};
\node[black] at (axis cs: 0.26,4) {(b)};

\addplot[thick,color=red!90!black,smooth] table [x index=0,y index=3,col sep=comma]
        {./img/jitter-comp-results-withVHT2.csv};
\node[red!90!black] at (axis cs: 0.45,26) {(c)};

\addplot[thick,color=green!60!black,smooth] table [x index=0,y index=4,col sep=comma]
        {./img/jitter-comp-results-withVHT2.csv};
\node[green!60!black] at (axis cs: 0.0045,89) {(d)};

\end{axis}

\end{tikzpicture}
 \end{center}
 \caption{Comparison of the jitter standard deviation in the WandStem node. Jitter of the 32768 Hz oscillator (a), 48MHz oscillator (b), jitter-compensated VHT (c) and original VHT (d). The light blue block is the measurement resolution.}
 \label{fig:jitter-comp-results}
\end{figure*}

The jitter-compensated VHT was implemented on the WandStem WSN node~\cite{bib:TerraneoEtAl-2016a} using the Miosix operating system~\cite{miosix}.
The entire timekeeper interface has been implemented, with the OS relying on the VHT \emph{get\_time} and \emph{set\_event} operations for task scheduling, and three hardware event sources:
\begin{enumerate}
\item the radio transceiver packet reception event, linked to an \emph{get\_hw\_event\_timestamp} operation;
\item the transceiver ``send packet'' line, controlled by a \emph{set\_hw\_event} operation;
\item an event line dedicated to applications, which can be configured as either a timestamp input
      through a \emph{get\_hw\_event\_timestamp}, or an event output via \emph{set\_hw\_event}.
\end{enumerate}

\begin{table}[h]
\centering
\caption{Number of capture/compare channels required to implement the timekeeper operations.}
\begin{tabular}{c|c|l}
jitter-compensated VHT & VHT & Operation\tabularnewline
\hline 
2 & 1 & Internal VHT requirements\tabularnewline
- & - & OS get\_time\tabularnewline
1 & 2 & OS set\_event\tabularnewline
1 & 2 & Radio get\_hw\_event\_timestamp\tabularnewline
1 & 2 & Radio set\_event\tabularnewline
1 & 2 & Application dedicated line\tabularnewline
\end{tabular}
\label{tab:hardwareRequirements}
\end{table}

Table~\ref{tab:hardwareRequirements} shows the hardware requirements in terms of capture/compare channels that were necessary to implement the timekeeper using the jitter-compensated VHT, compared to the requirements that would have been necessary to implement the original VHT algorithm. The first line refers to the requirements for the VHT to function not tied to specific operations, i.e., to the resources needed just to perform offset and skew correction in the jitter-compensated VHT, and to timestamp each rising edge of the slow clock in the original VHT. As can be seen, the jitter-compensated VHT requires 6 capture/compare channels, while the original VHT would have required 9 in the WandStem use case.

\subsection{Jitter attenuation}

Using the WandStem implementation, we measured the jitter of the original and jitter-compensated VHT, and compared it to measurements of the jitter of the fast and slow clock. Jitter measurements were performed by repeatedly measuring the length of an interval timed by the clock under test, and computing the standard deviation, except for the original VHT case in which the more precise clock is used to generate events timestamped by the VHT algorithm. The experiment was repeated for different time intervals to appreciate jitter integration over different time spans.

As measuring intervals of a clock requires another clock of greater precision, a FE5680 rubidium oscillator was used. The measurement resolution of our setup is 11.9 ns. 

Figure~\ref{fig:jitter-comp-results} shows the results. As can be seen, the 32768 Hz oscillator has a high jitter standard deviation, between 60 and 80 ns. The measure of the jitter of the 48 MHz turned out to be difficult due to being less than the measurement resolution. In these condition, the introduced quantization does not allow to make conclusions on the exact value, other than it is less than 11.9 ns.

The jitter-compensated VHT jitter is comparable to that of the 48 MHz oscillator for intervals up to 100ms, after which it starts increasing. This is due to the skew compensation controller, whose filtering action on jitter gets progressively less effective for longer periods.
Although this is unavoidable, and the performance of the jitter-compensated VHT will asymptotically -- and obviously -- reach that of the slow clock for longer periods, it can be argued that the effect of jitter is most relevant for small time interval measurements, where it can induce a high relative error. For larger intervals, jitter is less of concern compared to other error sources, such as thermal-induced skew variations over time.

The jitter of the original VHT is always equal or higher than the one of the 32768 Hz oscillator. Note that to achieve this results the timestamps affected by the original VHT race condition have been removed, otherwise its standard deviation would have been even higher.

\subsection{Power consumption}

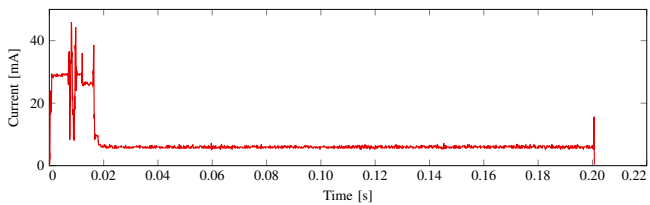
\begin{figure}[tb]
 \centerline{\resizebox{0.98\columnwidth}{!}{\begin{tikzpicture}
\pgfplotsset{width=0.90\textwidth,height=0.30\textwidth,
            compat=newest,
            major x grid style={very thin,gray},
            minor x grid style={very thin,dashed,gray},
            major y grid style={very thin,dashed,cyan!50!black},
}

\begin{axis}
[
xmin=0,
xmax=0.22,
xtick={0,0.02,0.04,0.06,0.08,0.10,0.12,0.14,0.16,0.18,0.20,0.22},
xticklabels={\rlap{0},0.02,0.04,0.06,0.08,0.10,0.12,0.14,0.16,0.18,0.20,\llap{0.22}},
ymin=0,
ymax=50,
enlargelimits=false,
xlabel={Time [s]},
ylabel={Current [mA]},
]
\addplot[thick,color=red!90!black,smooth] table [x index=0,y index=1,col sep=comma]
        {./img/flopsyncvht-power-trace.csv};

\end{axis}

\end{tikzpicture}}}
 \caption{Power trace of a WandStem node waking up from deep sleep to perform inter-node clock synchronization using FLOPSYNC-2, propagation delay compensation and intra-node jitter-compensated VHT synchronization.}
 \label{fig:flopsyncvht-power-trace}
\end{figure}

To assess the current consumption of the jitter-compensated VHT we performed a test where a WandStem node performs inter-node clock synchronization using FLOPSYNC-2 and propagation delay compensation, using jitter-compensated VHT as the underlying timebase.

Figure~\ref{fig:flopsyncvht-power-trace} shows a trace of the node current consumption. Upon wakeup from deep sleep, the node first resynchronizes the fast clock to the slow one as shown in Section~\ref{sec:naiveSolution}. Then it waits for the (inter-node) synchronization packet. The two peaks above 40mA are the Glossy~\cite{bib:FerrariEtAl-2011a} rebroadcast of the synchronization packet, and the packet sent to estimate propagation delays. The node then waits for the propagation delay reply and enters the sleep state with the high frequency clock active until 200ms have passed, to estimate the skew as explained in Section~\ref{sec:skewCorrection}. Finally, the node can go to deep sleep till the next sync period.

Notice that despite the requirement for the jitter-compensated VHT to spend 200ms to measure the relative skew, the node can still spend 98.0\% (10s sync period) and 99.7\% (60s sync period) in the deep sleep state.

To estimate the power saving of the jitter-compensated VHT, the experiment of Figure~\ref{fig:flopsyncvht-power-trace} has been repeated with no VHT, thus with only the high frequency clock and no deep sleep. To compare the jitter-compensated VHT with the original one, we estimated its consumption by running jitter-compensated VHT with intra-node skew compensation disabled, which is what sets it apart in terms of power consumption. We could not use the original VHT implementation of Section~\ref{sec:vhtIssues} because it is incapable of going in deep sleep, but this does not affect the purpose of the test. Results are summarized in Table~\ref{tab:consumptionComparison}

\begin{table}[h]
\centering
\caption{Average synchronization consumption comparison.}
\begin{tabular}{c|rr}
Sync period & 10s & 60s\tabularnewline
\hline
Without VHT  &  6035uA & 6006uA\tabularnewline
Flopsync VHT &   158uA &   28uA\tabularnewline
VHT          &    30uA &    7uA\tabularnewline
\end{tabular}
\label{tab:consumptionComparison}
\end{table}

\section{Conclusions and future work}
\label{sec:Conclusions}

This paper focused on one of the key methodologies that allowed joint low power and high resolution timekeeping in WSNs, namely Virtual High-resolution Time.
Its operation was studied, identifying three issues as jitter susceptibility, high hardware requirements, and a race condition.

A new solution was proposed, which fully casts the VHT in the framework of clock synchronization. The proposed solution was implemented on a real WSN node, assessing the expected performance improvement. At present the only relevant drawback of the proposed solution is an increased computation overhead, still compatible with the applications, but worth further research effort.

It is expected that the research here presented will foster a more widespread adoption of VHT timebases, by lowering the barrier required for their implementation, and making them more palatable thanks to the improved accuracy.

Furher work will also address the use of the proposed technique to tackle ``blended skews'' also in the context of inter-node clock synchronization problem.%


\end{document}